\documentclass[conference,letterpaper]{IEEEtran}

\usepackage[utf8]{inputenc} 
\usepackage[T1]{fontenc}
\usepackage{url}
\usepackage{ifthen}
\usepackage{comment}
\usepackage{graphicx}
\usepackage{cite}
\usepackage{amssymb}
\usepackage[nolist]{acronym}
\usepackage[cmex10]{amsmath} 

\newtheorem{theorem}{Theorem}

\newtheorem{proof}{Proof}

\newtheorem{remark}{Remark}

\begin{acronym}
    \acro{GDI}{General Definition of Information}
    \acro{IRP}{Inverse Relationship Principle}
    \acro{AEP}{Asymptotic Equipartition Property}
    \acro{p.m.f.}{probability mass function}
    \acro{LoA}{Levels of Abstraction}
    \acro{i.i.d.}{independent and identically distributed}
\end{acronym}
\begin{document}
\title{Semantic Communication: From Philosophical Conceptions Towards a Mathematical Framework} 


\author{%
  \IEEEauthorblockN{Javad Gholipour, Rafael F. Schaefer, and Gerhard P. Fettweis}
  \IEEEauthorblockA{
  Technische Universität Dresden, Dresden, Germany\\
  Email: javad.gholipour@tu-dresden.de, rafael.schaefer@tu-dresden.de, gerhard.fettweis@tu-dresden.de}
}

\maketitle

\begin{abstract}
Semantic communication has emerged as a promising paradigm to address the challenges of next-generation communication networks. While some progress has been made in its conceptualization, fundamental questions remain unresolved. In this paper, we propose a probabilistic model for semantic communication that, unlike prior works primarily rooted in intuitions from human language, is grounded in a rigorous philosophical conception of information and its relationship with data as Constraining Affordances, mediated by \ac{LoA}. This foundation not only enables the modeling of linguistic semantic communication but also provides a domain-independent definition of semantic content, extending its applicability beyond linguistic contexts. As the semantic communication problem involves a complex interplay of various factors, making it difficult to tackle in its entirety, we propose to orthogonalize it by classifying it into simpler sub-problems and approach the general problem step by step. Notably, we show that Shannon's framework constitutes a special case of semantic communication, in which each message conveys a single, unambiguous meaning. Consequently, the capacity in Shannon's model—defined as the maximum rate of reliably transmissible messages—coincides with the semantic capacity under this constrained scenario. In this paper, we specifically focus on the sub-problem where semantic ambiguity arises solely from physical channel noise and derive a lower bound for its \textbf{semantic} capacity, which reduces to Shannon's capacity in the corresponding special case. We also demonstrate that the achievable rate of all transmissible messages for reliable semantic communication, exceeds Shannon's capacity by the added term $H(X|S)$.
\end{abstract}


\section{Introduction}
Information has long been one of the most challenging and intriguing concepts, attracting the attention of philosophers for centuries. Despite significant efforts, philosophical studies on (semantic) information remain underdeveloped. In the past century, particularly following Shannon's pioneering work \cite{Shannon}, communication engineers have also become deeply engaged with this concept. Shannon defined information and quantified it using a measure he called entropy. Based on that, he established the fundamental limits of source compression and reliable transmission over noisy channels. Shannon's theory revolutionized communication systems, laying the groundwork for the remarkable advancements in communication technologies and networks that we see today.

One year after Shannon's groundbreaking work \cite{Shannon}, Weaver categorized communication problems into three distinct levels: technical, semantic, and effectiveness \cite{Weaver}. The technical level addresses the question, "How accurately can the symbols of communication be transmitted?" The semantic level focuses on, "How precisely do the transmitted symbols convey the intended meaning?" Finally, the effectiveness level examines, "How effectively does the received meaning influence behavior as desired?" Weaver argued that Shannon's theory primarily addresses the technical level. However, he also noted that Shannon's framework is sufficiently general to provide a foundation for modeling the semantic and effectiveness levels by incorporating relevant elements.

Despite the challenges and the lack of a clear and widely-accepted (philosophical) definition of (semantic) information, its data-based interpretation has gained significant attention over the past few decades \cite{Floridi:1}. Although the concept of data itself remains philosophically underexplored, the bipartite definition of information as "data + meaning", known as the \ac{GDI}, has been widely accepted by many philosophers and engineers, becoming an operational standard in many disciplines \cite{Sommaruga}. From the \ac{GDI} perspective, information understood as semantic content must consist of well-formed and meaningful data \cite{Sommaruga}. "Well-formed" describes data that is properly organized according to the syntax—defined as the rules that govern a specific system, code, or language. 
"Meaningful" signifies that the data adheres to the semantics, or meanings, relevant to the chosen system, whether conveyed visually, linguistically, or in other forms. 

Shannon's theory became widely known as "information theory." While he reportedly came to regret the popularity of this term, most researchers studying the concept of information agree that his theory established a rigorous framework that constrains further theorization \cite{Sommaruga}. Two key links between Shannon's theory and the philosophy of (semantic) information have remained stable: the communication model, which has been largely unchallenged, and the \ac{IRP}, which highlights the inverse relationship between the probability of an event and the amount of semantic information it conveys \cite{Sommaruga}.

Bar-Hillel and Carnap \cite{Carnap} were among the first to attempt quantifying semantic information. Utilizing what they termed "logical probability," they measured the semantic content of a sentence and referred to this measure as semantic entropy. Their approach focused exclusively on propositional logic, defining the logical probability of a sentence as the likelihood of its truth across all possible scenarios. However, their theory had several limitations. First, it was confined to propositional logic and lacked generality. Second, it paradoxically assigned maximal information to contradictory sentences. To address the latter, Floridi \cite{Floridi:2} incorporated the truthfulness into their framework. While this resolved the problems, the approach still lacked generality because it needs a reference for truthfulness and is limited to the propositional logic as well. Many attempts have been made to quantify semantic information, but each is tailored to its own specific problem domain \cite{Gunduz}. 

Noise is a critical factor limiting (semantic) communication, making its study vital for theoretical advancements. While physical channel noise in classical communication is well understood, semantic communication introduces two types of noises: physical channel noise and semantic channel noise. The latter arises from mismatched knowledge backgrounds between sender and receiver, causing misunderstandings even when the transmitted signal is received correctly. For instance, physical channel noise might alter a transmitted word (``cat'' to ``bat''), changing its meaning. Conversely, semantic channel noise could cause misinterpretation, such as the word ``football''  being understood differently in the U.S. and U.K. Researchers have defined semantic noise in various ways: Qin et al. \cite{Qin} described it as disturbances impairing message interpretation, Shi et al. \cite{Shi} highlighted it as errors causing misunderstandings, and Hu et al. \cite{Hu} emphasized misalignments between intended and received semantic symbols.

Bao et al. proposed a semantic communication model extending Shannon's framework to include semantic elements, using Bar-Hillel and Carnap's semantic entropy to quantify semantic information and developing theories for semantic source and channel coding, though limited to propositional logic \cite{Bao, Carnap}. Ma et al. \cite{Ma} defined semantics as a subset of the message set, encoding and reconstructing it at the receiver to derive semantic channel capacity. However we argue that, natural semantic communication involves first generating semantics, then expressing them through messages, and finally recovering them from received messages. 


Shao et al. \cite{Shao} propose a probabilistic model for semantic communication, grounded in illustrative examples from human language. Their central observation—that a single message can convey multiple meanings, and a single meaning can be represented through various messages—highlights key semantic ambiguities in linguistic communication. However, their framework lacks grounding in foundational theories of semantic information, such as the \ac{GDI} \cite{Sommaruga}, which forms the theoretical basis of our model. As a result, their approach remains constrained to the domain of natural language, without a clear path toward generalization across diverse data types and representational forms, and may even lack formal validity within linguistic theory itself, as it is based solely on observational examples rather than established linguistic principles.

In contrast, we present a probabilistic model of semantic communication that is explicitly rooted in a rigorous philosophical conception of information, inspired by Weaver’s extension of Shannon’s model \cite{Weaver}, and formally grounded in the \ac{GDI} framework \cite{Floridi:1,Sommaruga}. By incorporating core notions such as well-formed data, meaningfulness, and the independence of semantic content from its physical support, our model extends beyond linguistic semantics to encompass information in any representational modality. As Floridi notes, ``\textit{The dependence of information on the occurrence of syntactically well-formed data, and of data on the occurrence of differences variously implementable physically, explain why information can so easily be decoupled from its support. The actual format, medium and language in which semantic information is encoded is often irrelevant and hence disregardable. In particular, the same semantic information may be analog or digital, printed on paper or viewed on a screen, in English or in some other language, expressed in words or pictures.}'' \cite{Sommaruga}. 

While our model arrives at a similar observation regarding the dual mappings between messages and meanings, it does so through a more general and formally defensible lens. To develop a systematic theory, we decompose the semantic communication problem into multiple sub-problems. In this paper, as the first step towards a general theory, we address the foundational case in which semantic noise arises solely from physical channel imperfections. We demonstrate that Shannon’s classical model emerges as an extreme case of our broader semantic framework, and we derive an achievable semantic communication rate that reduces to Shannon capacity in this case—while the achievable rate of all transmissible messages exceeding it by the added term $H(X|S)$.

\begin{figure}
\vspace{-\baselineskip}
\centering
\includegraphics[width=0.50\textwidth]{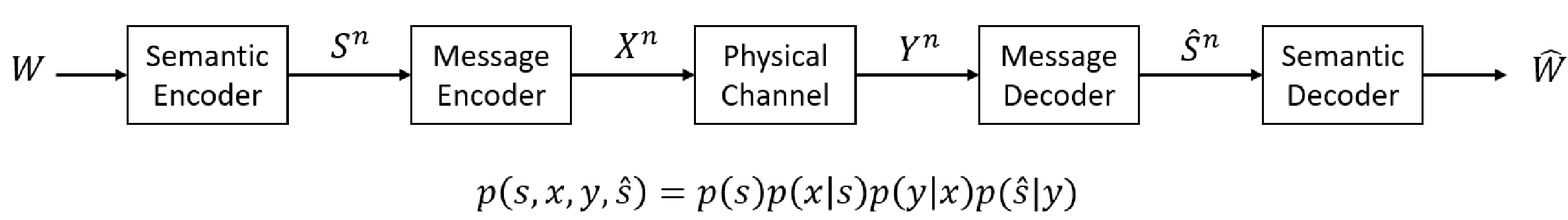}
\caption{The discrete memoryless semantic communication channel.}
\label{Fig1}
\vspace{-\baselineskip}
\end{figure}

\textit{Notation}: we use capital letters to show random variables (e.g., $X$); lower case letters for realization of random variables (e.g., $x$) and $p(x)$ to show the \ac{p.m.f.} of random variable $X$ on set $\mathcal{X}$. We also define $X^n$ as an $n$-sequence. Also the set of $\epsilon$-strongly joint-typical of $n$-sequences $X^n$ and $Y^n$ with \ac{p.m.f.} $p(x,y)$, is shown by $T_\epsilon^n(X,Y)$. The point-to-point discrete memoryless semantic communication channel, as shown in Figure \ref{Fig1}, is defined with the alphabet sets ($\mathcal{S},\mathcal{X},\mathcal{Y},\hat{\mathcal{S}}$) and a probability transition function $p(\hat{s},y|s,x)$ defined for all $(s,x,y,\hat{s})\in \mathcal{S}\times\mathcal{X}\times\mathcal{Y}\times\hat{\mathcal{S}}$, where $S^n,X^n$ are the channel inputs and $\hat{\mathcal{S}}^n,Y^n$ are the channel outputs. The transmitter wishes to transmit a semantic $w=1$ which is uniformly distributed over the semantic set $\mathcal{W}$ with cardinality $|\mathcal{W}|$. We define $\frac{\log |\mathcal{W}|}{n}$ as the semantic rate.

\section{Probabilistic Model of Semantic Communication}

In this section, we develop a probabilistic model for semantic communication based on the \ac{GDI} components—data and information—and their interrelationship. We show that Shannon's model, which addresses the technical aspects of communication \cite{Weaver}, is a special case of our model. We also analyze semantic noise \cite{Shi}, consisting of semantic channel noise (due to mismatched background knowledge) and physical channel noise, and provide an information theoretic explanation of semantic channel noise, leveraging Shannon's model as an extreme case.

\begin{figure}
\vspace{-\baselineskip}
\centering
\includegraphics[width=0.50\textwidth]{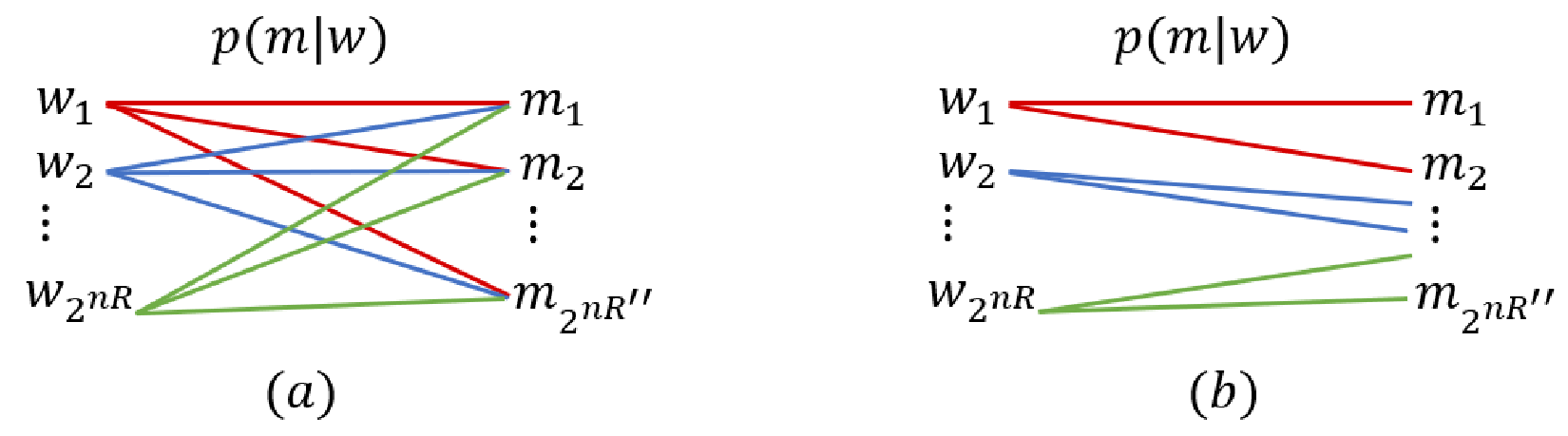}
\caption{Interrelationship between semantics $W$ and the messages $M$, (a): in general, and (b): for a given context $Q$.}
\label{Fig2}
\vspace{-\baselineskip}
\end{figure}

As outlined in \ac{GDI}, (semantic) information comprises well-formed and meaningful data. The connection between well-formed data and meaningfulness is mediated by two critical concepts: Data as Constraining Affordances and \ac{LoA} \cite{Sommaruga}. Data as Constraining Affordances refers to the idea that data constituting (factual) information invites certain interpretations while resisting others. For instance, the word bank most commonly refers to a financial institution in everyday contexts, whereas its meaning as the side of a river is less likely and typically arises only within specific environmental or linguistic settings. Importantly, this concept is not limited to human language but extends to various forms of data and information across different domains. For example, in cellular signaling in biology, a single message such as insulin secretion can be interpreted differently depending on the cell type: it may trigger glucose uptake in muscle cells, promote fat storage in adipocytes, or inhibit glucose production in the liver. Similarly, in neural encoding within the brain, a spike train in the visual cortex might represent the perception of the color red under normal conditions, or it could represent a memory of the color red without visual input, or even a different object perceived as red, depending on the context. In environmental monitoring systems, a sensor spike indicating infrared heat could be interpreted as a fire, animal movement, or even a false positive due to environmental factors. These examples illustrate how data as constraining affordances provides a generalizable framework for interpreting data—not only within linguistic systems but also in biological, cognitive, technological environments, etc. In all these examples, it is illustrated that data affords multiple interpretations, each constrained by the observer’s context. 

Based on the concept of data as constraining affordances, we can model the relationship between a message (data) and its possible semantics (interpretations) probabilistically, in a general way that is not limited only to linguistic communication. Let $W$ and $M$ represent random variables corresponding to semantics and messages, respectively. Each message $m$ can be interpreted by multiple semantics $w$ with some probability $p(w|m)$. Similarly, each semantic $w$ can be expressed by various messages $m$ with some probability $p(m|w)$, as illustrated in Figure \ref{Fig2}(a). This model highlights that, given a message $m$, there is inherent semantic ambiguity at the information source, as $m$ can correspond to multiple semantics $w$. However, once a message is produced, the information source itself knows the intended semantic $w$. For example, as humans, we inherently grasp the meaning behind a sentence, facial expression, or body language we convey—even though these may carry different interpretations across varying contexts. To capture this
, additional modeling is required. 

\begin{figure}
\vspace{-\baselineskip}
\centering
\includegraphics[width=0.40\textwidth]{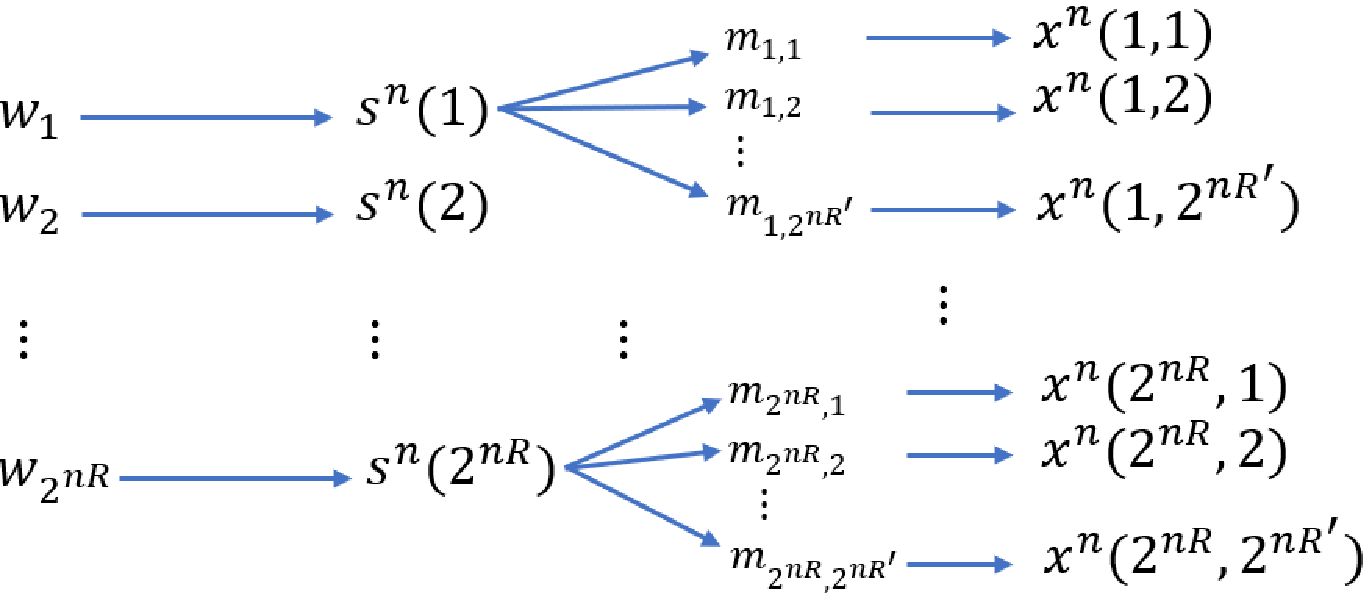}
\caption{General semantic-message encoder given the context $Q$.}
\label{Fig31}
\vspace{-\baselineskip}
\end{figure}

\ac{LoA} provide a framework for the mentioned purpose. \ac{LoA}
act as bridges between an information agent (observer) and their physical or conceptual environment, influencing how information is interpreted or utilized \cite{Sommaruga}. For example, the phrase "Explain it again" might lead to different interpretations: at a teacher's \ac{LoA}, it might be elaborated as a request for a detailed explanation, while at a student's \ac{LoA}, it might be viewed as a call for a brief summary. To model and generalize the \ac{LoA}, we propose another factor—context $Q$—which represents the sufficient information that ensures each message $m$ corresponds to a single, unambiguous semantic $w$. To clarify the role of context, consider human communication, where factors such as location, time, psychological state, and numerous other parameters ensure that each expressed message conveys a singular, specific meaning. As shown in Figure \ref{Fig2}(b), the inclusion of context enables the partitioning of the set of all possible messages into subsets, where each subset corresponds to only one specific semantic.
Mathematically, we can represent the relationships between semantics $W$, messages $M$, and context $Q$ as follows
\begin{align}
    &H(W|M)>0,\\
    &H(W|M,Q)=0.
\end{align}

In summary, given the context $Q$, we propose to model the semantic-message encoder as illustrated in Figure \ref{Fig31}. Consider $2^{nR}$ semantics $W$, represented by $w_{1}, w_{2}, \ldots, w_{2^{nR}}$. Each semantic $W$ is mapped to a sequence $S^n$, represented by $s^n(1), s^n(2), \ldots, s^n(2^{nR})$. Assume there exist $2^{n(R+R')}$ possible messages $M$, which can express all $2^{nR}$ semantics. The set of messages can be partitioned into $2^{nR}$ equal-sized subsets, each containing $2^{nR'}$ messages. Each subset is interpretable by only one semantic $W$. For a different context $Q$, this partitioning changes, meaning that for different contexts, some messages may correspond to different meanings. In total, there exist $\binom{2^{n(R+R')}}{2^{nR'}, 2^{nR'}, \ldots, 2^{nR'}}$ distinct ways to partition the $2^{n(R+R')}$ messages into $2^{nR}$ equal-sized subsets, such that $2^{nR'} + 2^{nR'} + \ldots + 2^{nR'} = 2^{n(R+R')}$. Therefore, given the context $Q$, each encoded semantic $s^n(i)$ can be expressed by $2^{nR'}$ messages $M$, represented as $m_{i,1}, m_{i,2}, \ldots, m_{i,2^{nR'}}$. To transmit the encoded semantic $s^n(i)$ over the physical channel, it must first be processed by the channel encoder. Each semantic code $s^n(i)$ can be transmitted through the physical channel using any of the corresponding channel codes $x^n(i,1), x^n(i,2), \ldots, x^n(i,2^{nR'})$, which are associated with the expressible messages $m_{i,1}, m_{i,2}, \ldots, m_{i,2^{nR'}}$.

\begin{figure}
\vspace{-\baselineskip}
\centering
\includegraphics[width=0.35\textwidth]{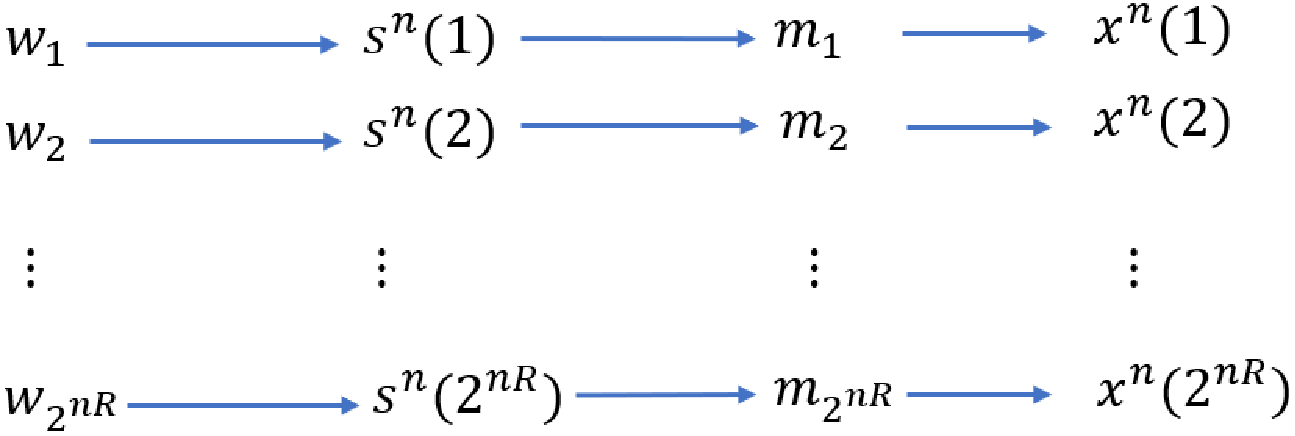}
\caption{Semantic-message encoder given the context $Q$, for the extreme case, where the semantic set $\mathcal{W}$ and the message set $\mathcal{M}$ have one-to-one relation.}
\label{Fig32}
\vspace{-\baselineskip}
\end{figure}

\begin{remark}
    Consider the extreme case where each semantic $W$ is expressible by exactly one message $M$, and each message $M$ is interpretable by only one semantic $W$,
    as depicted in Figure \ref{Fig32}. This case resembles Shannon's model, where each message has a unique meaning. Consequently, in Shannon's framework, the left half of the Figure \ref{Fig32}, which models the semantic layer, is irrelevant, and the focus shifts entirely to the codebook that relates the message set $\mathcal{M}$ to the channel codes $\mathcal{X}^n$. This shows that Shannon's model is an extreme case of our proposed semantic communication model.
\end{remark}

As discussed in the previous section, semantic ambiguity can arise from both semantic channel noise and physical channel noise. 
Semantic channel noise has been defined in the literature as the discrepancy in the knowledge background between the sender and the receiver \cite{Shi}. To better understand and frame the semantic channel noise, we focus on the extreme case of our proposed model, which showed to be Shannon's model (Figure \ref{Fig32}).
One of the most important assumptions in Shannon's work to prove his channel coding theorem was that he fully declared the codebook (as the mapping between the set of messages $\mathcal{M}$ and the channel codes $\mathcal{X}^n$ generated by the sender) to the receiver. What happens if the codebook is not fully shared between the sender and receiver? 

\begin{figure}[htp]
\centering
\includegraphics[width=0.35\textwidth]{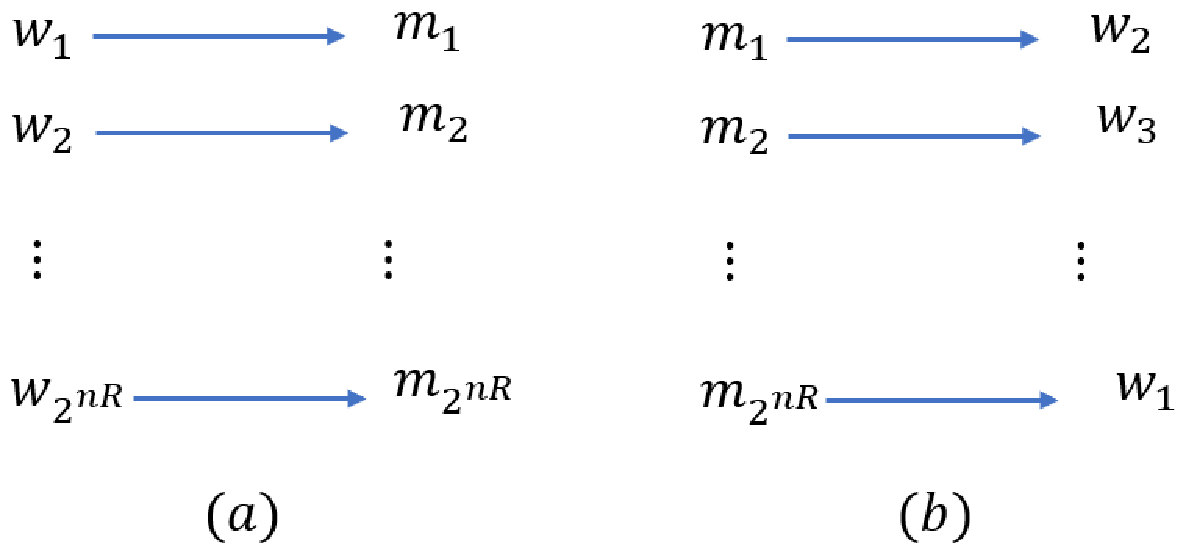}
\caption{Semantic channel noise in Shannon's model, an extreme case of semantic communication, illustrated as mismatched codebooks between the sender (a) and receiver (b).}
\label{Fig4}
\end{figure}

\begin{figure}
\vspace{-\baselineskip}
\centering
\includegraphics[width=0.4\textwidth]{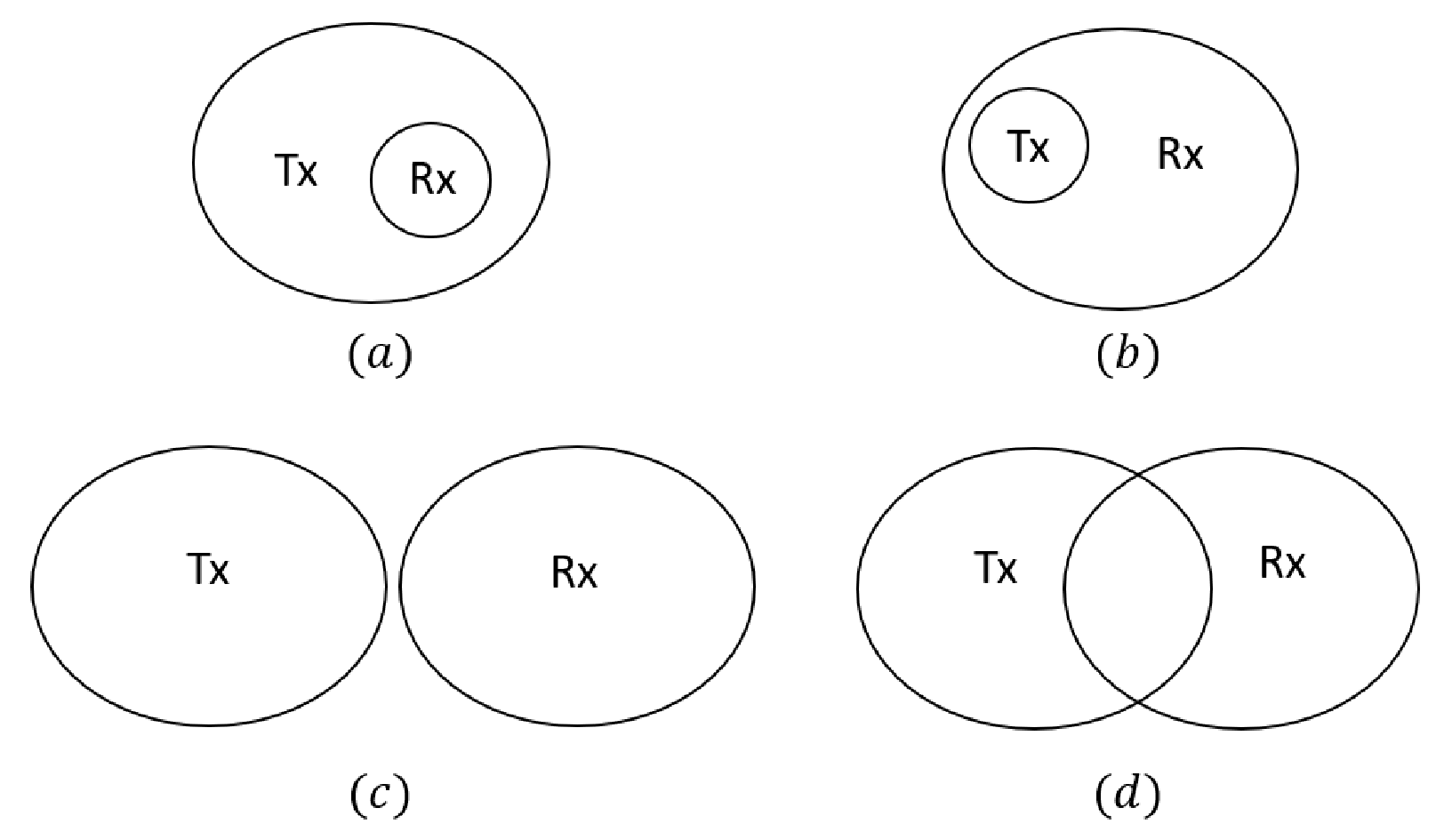}
\caption{Scenarios of sender's and receiver's background knowledges: (a) sender knows receiver's knowledge, (b) receiver knows sender's knowledge, (c) no shared knowledge, (d) some shared knowledge.}
\label{Fig5}
\vspace{-\baselineskip}
\end{figure}

In Shannon's model, assume the transmitter's codebook is as shown in Figure \ref{Fig4}(a), and the receiver is provided with its shifted version, as in Figure \ref{Fig4}(b). While the message may be decoded correctly, its interpretation could fail due to semantic channel noise, modeled by the disparity between the sender's and receiver's codebooks. By proper re-shifting the codebook on either side, we can eliminate this noise and avoid semantic ambiguity. Semantic channel noise can be classified into four cases: (1) the receiver’s knowledge is a subset of the sender’s; (2) the sender’s knowledge is a subset of the receiver’s; (3) their knowledge is disjoint, causing ambiguity; and (4) their knowledge intersects fully (Shannon’s case) or partially. in these cases, illustrated in Figure \ref{Fig5}, the semantic ambiguity can be resolved through adaptive codebook design, where adjusting the codebooks ensures correct message interpretation, as in cases (1) and (2), a mother adapts her communication to match the child's understanding.

Semantic noise can result from both semantic and physical channel noise. In this paper, we focus on the case where semantic noise is caused solely by physical channel noise, with no semantic channel noise. This implies that the codebook is fully shared with the receiver, eliminating ambiguity in message interpretation due to background knowledge disparities.

\section{Main Results}

In this paper, we examine the semantic communication problem for the case where the context is assumed to be fixed, e.g., $Q=q$, and the semantic ambiguity arises solely from physical channel noise, with the semantic channel assumed to be noiseless—i.e., the generated codebook is fully shared between the transmitter and receiver. For simplicity, we do not explicitly include $Q$ in our mathematical expressions.

Consider the point-to-point semantic communication system model depicted in Figure \ref{Fig1}, where the sender wishes to reliably transmit its intended meaning (semantic) $W$ over a communication channel consisting only the physical channel noise. A $(2^{nR},2^{nR'},n)$ code for the discrete memoryless semantic communication channel consists of:
\begin{enumerate}
  \item A semantic set $\mathcal{W}=\{1,...,2^{nR}\}$,
  \item A semantic encoding function $f_1:\mathcal{W}\rightarrow{\mathcal{S}}^n$,
  \item A channel encoding function $f_2:{\mathcal{S}}^n\rightarrow{\mathcal{X}}^n$,
  \item A decoding function $g:{\mathcal{Y}}^n\rightarrow\mathcal{W}$.
\end{enumerate}

The average probability of error can be expressed as:
\begin{align}
    P_e^{(n)} = 2^{-nR}\sum_{w=1}^{2^{nR}}P(g(Y^n)\neq w | w~\text{is sent}).
\end{align}

\begin{figure}
\vspace{-\baselineskip}
\centering
\includegraphics[width=0.35\textwidth]{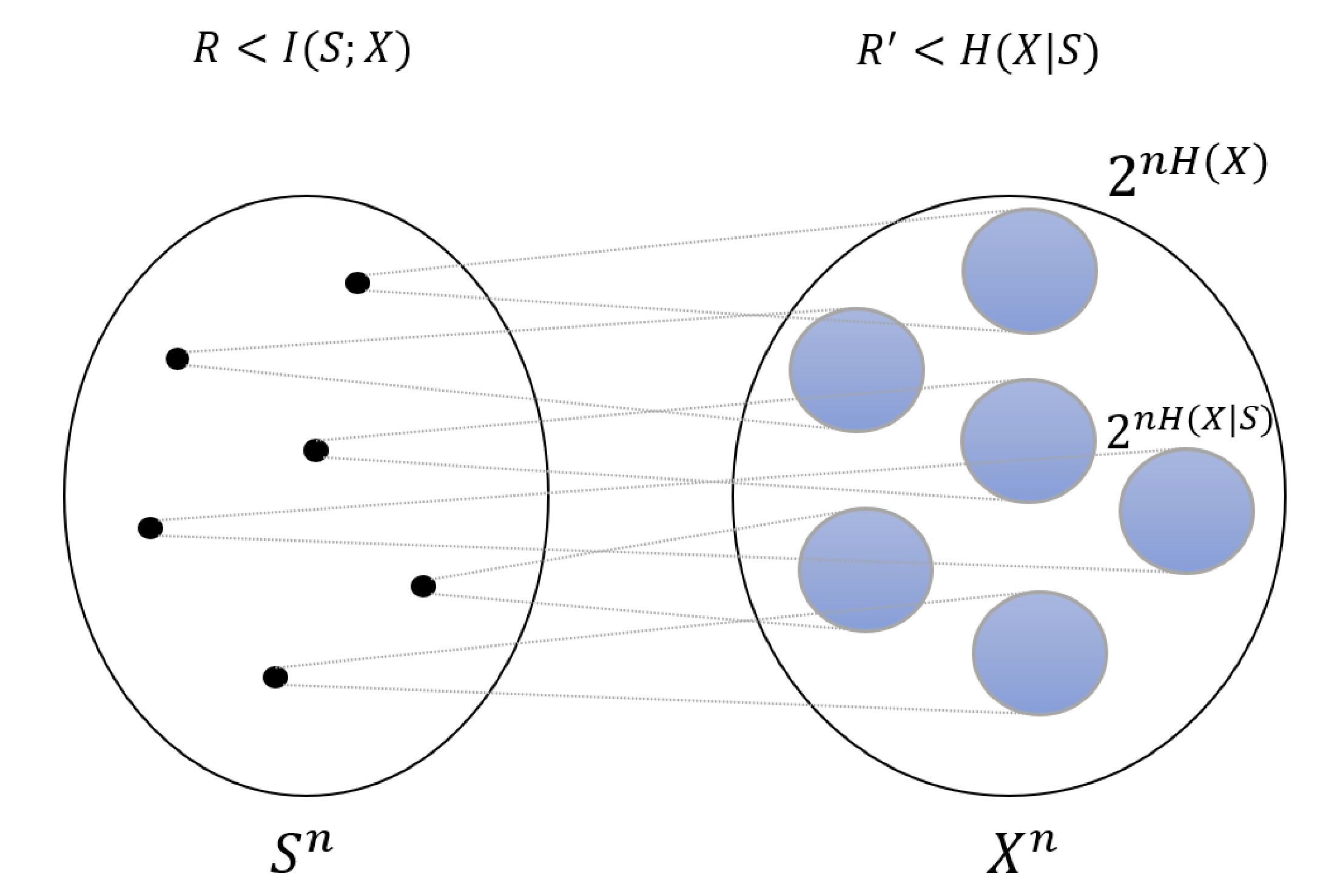}
\caption{Encoding for semantic communication.}
\label{Fig6}
\vspace{-\baselineskip}
\end{figure}

\begin{theorem}\label{th:1}
    For the discrete memoryless semantic communication channel (Figure \ref{Fig1}), where the semantic noise is only caused by the physical channel noise and the semantic channel is noiseless, the following rate region is achievable:
    \begin{align}\label{rates}
        R\leq I(S;Y),\\\nonumber
        R'\leq H(X|S),
    \end{align} 
    where $R$ represents the semantic rate and $R'$ denotes the message rate expressing each semantic.
    
\end{theorem}
\begin{remark}\label{rem:1}
    In Theorem \ref{th:1}, the rate of all transmissible messages $R+R'$ must satisfy the following condition:
    \begin{align}\label{rem:11}
        R+R'\leq H(X|S)+I(S;Y)\leq H(X|S)+I(X;Y),
    \end{align}
    where the last inequality is due to the data processing inequality\cite{Cover}. This demonstrates that reliable semantic communication achieves an improved message rate compared to Shannon's rate, with an additional term of $H(X|S)$.
\end{remark}

\begin{remark}\label{rem:2}
    Theorem \ref{th:1} demonstrates that for reliable semantic communication, both the semantic rate and the expressibility of the source through messages are essential factors.
\end{remark}

\begin{remark}\label{rem:3}
    In Theorem \ref{th:1}, setting $R'=0$, results in $H(X|S)=0$. In this scenario, a one-to-one relationship exists between semantics and messages (Shannon's model as extreme case of our semantic communication model), and the rate of all transmissible messages in (\ref{rem:11}) simplifies to $R+R'=R\leq I(X;Y)$. 
    We demonstrate that, in this case, the achievable semantic rate and the rate of all transmissible messages as expected are equivalent and both equal to Shannon's capacity.
\end{remark}


Before proving the Theorem \ref{th:1}, we provide an intuitive discussion. As outlined in the previous section, given the context $Q$ there must be no semantic ambiguity at the transmitter. This implies that each message $M$, encoded by $X^n$, must correspond to a unique semantic $W$, encoded semantically by $S^n$. As illustrated in Figure \ref{Fig6}, this requires that the blue spheres on the right-hand side do not overlap, which imposes the condition that the semantic rate must satisfy $R\leq I(S;X)$. Additionally, it is evident that the rate of the messages $R'$ expressing each semantic, must satisfy $R'\leq H(X|S)$.

To transmit the semantic $W$ over the physically noisy channel, one of the associated codewords $X^n$, corresponding to the expressing messages $M$ of the selected semantic $w$ (associated with $S^n$), is chosen uniformly at random and transmitted through the channel. At the receiver side as depicted in Figure \ref{Fig7}, it is not necessary to decode the exact message $M$ associated with the transmitted codeword $X^n$. Instead, it suffices to decode any of the messages $M$ corresponding to the selected semantic $W$ at the transmitter. This means that on the right sphere in Figure \ref{Fig7}, the blue spheres corresponding to the semantics $W$ must not overlap. However, the smaller spheres within each blue sphere, which correspond to the messages $M$ expressing the associated semantic $W$, are allowed to overlap. Consequently, the rate of expressing messages for each semantic, $R'$ must satisfy $R'\geq I(X;Y|S)$ and the total rate of all transmissible messages must satisfy $R+R'\leq I(S,X;Y)$. 

\begin{proof}
Now we start the proof of Theorem \ref{th:1} as follows:
\subsubsection{Codebook Generation}
Consider $\binom{2^{n(R+R')}}{2^{nR'},2^{nR'},...,2^{nR'}}$ contexts $Q$, where $2^{nR'}+2^{nR'}+...+2^{nR'}=2^{n(R+R')}$. Consider the message set $\mathcal{M}=\{1,...,2^{n(R+R')}\}$, for which, there are $\binom{2^{n(R+R')}}{2^{nR'},2^{nR'},...,2^{nR'}}$ distinct way to partition it into equal-sized subsets, each of cardinality $2^{nR'}$, such that the sum of the sizes of the subsets equals $2^{n(R+R')}$. Consequently, there are $2^{nR}$ distinct subsets. Let the semantic $W$ be uniformly distributed over the set $\mathcal{W}=\{1,...,2^{nR}\}$. Generate $2^{nR}$ \ac{i.i.d.} semantic codewords $S^n$ according to the \ac{p.m.f.} $p(s)$. For each semantic codeword $S^n$, generate $2^{nR'}$ \ac{i.i.d.} codewords $X^n$ according to the conditional \ac{p.m.f.} $p(x|s)$. So, there are a total of $2^{n(R+R')}$ codewords $X^n$.

\subsubsection{Encoding}
Associate each context $Q$ with one of the partitioning way that partitioned the message set $\mathcal{M}$. Fix the context such that $Q=q$, and partition the message set $\mathcal{M}$ with the associated partitioning way. Associate each semantic $w$ with a semantic codeword $S^n(w)$. Then associate each semantic codeword $S^n(w)$ with one of the subsets of $\mathcal{M}$. Furthermore, associate each semantic codeword $S^n(w)$ with $2^{nR'}$ codewords $X^n(w,m)$, labeled by the semantic $w$, and the messages $m\in \{1,...,2^{nR'}\}$ corresponding to the elements of the associated subset of $\mathcal{M}$.

To transmit the semantic $w$, the sender selects one of the codewords $X^n(w,m)$ uniformly at random from the $2^{nR'}$ available choices and transmits it over the physical channel.

\begin{figure}
\vspace{-\baselineskip}
\centering
\includegraphics[width=0.50\textwidth]{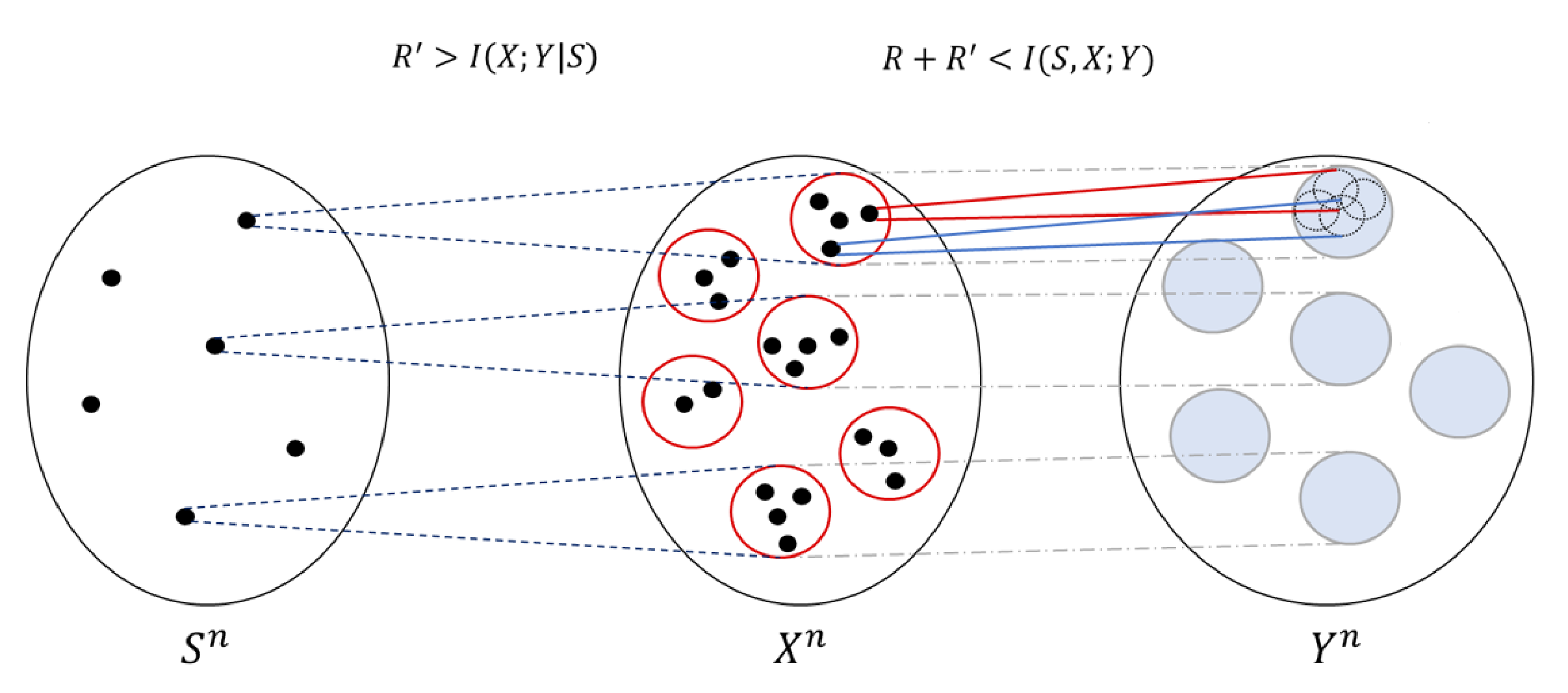}
\caption{Decoding for semantic communication.}
\label{Fig7}
\vspace{-\baselineskip}
\end{figure}

\subsubsection{Decoding}
The context $Q=q$ is known at the transmitter, ensuring there is no ambiguity on its side. This implies that each message must have a unique meaning. Mathematically, the probability of error at the transmitter is given by
\begin{align}
    P_{e,t}^{(n)} = 2^{-nR}\sum_{w=1}^{2^{nR}}P(f_1^{-1}(f_2^{-1}(X^n))\neq w | w~\text{is chosen})
\end{align}

Assume $w=1$. Therefore
\begin{align}\label{error}
    P_{e,t}^{(n)}\leq P(E_0)+P(E_1),
\end{align}
where
\begin{align}
    &E_0=\{(s^n(1),x^n(1,m))\notin T_\epsilon^n(S,X)~\text{for some}~m\},\\
    &E_1=\{(s^n(w),x^n(1,m))\in T_\epsilon^n(S,X)~\text{for some}~w\neq 1\}.
\end{align}

The first term on the right side of (\ref{error}) vanishes as $n\rightarrow\infty$ by the \ac{AEP} theorem. According to the joint typicality lemma \cite{Cover}, the second term also vanishes as $n\rightarrow\infty$, provided that:
\begin{align}\label{1}
    R\leq I(S;X).
\end{align}
\begin{remark}
    It is also worth noting that the size of the conditionally typical set is clearly upper bounded as:
    \begin{align}\label{2}
        |T_\epsilon^n(X|S)|\leq 2^{n(H(X|S))},
    \end{align}
    which implies that the rate of the message subset satisfies:
    \begin{align}\label{2}
        R'\leq H(X|S).
    \end{align}
\end{remark}

Declare the context $Q=q$ and the codebook to the receiver to mitigate errors caused by semantic channel noise. Assume $m=1$. Thus, the transmitter sends $x^n(1,1)$ through the channel. Upon receiving the sequence $y^n$, the receiver searches for the unique semantic $w$ such that:
\begin{align}
    \{(s^n(w),x^n(w,m),y^n)\in T_\epsilon^n(S,X,Y)~\text{for some}~m\}.
\end{align}

Therefore the probability of error is bounded as
\begin{align}\label{error:r}
    P_e^{(n)}\leq P(E_2)+P(E_3)+P(E_4)+P(E_5),
\end{align}
where
\begin{align}
    &E_2=\{(s^n(1),x^n(1,m),y^n)\notin T_\epsilon^n(S,X,Y)~\text{for all}~m\},\\
    &E_3=\{(s^n(1),x^n(w,m),y^n)\in T_\epsilon^n(S,X,Y)~\text{for some}~w\neq 1\},\\
    &E_4=\{(s^n(w),x^n(1,m),y^n)\in T_\epsilon^n(S,X,Y)~\text{for some}~w\neq 1\},\\
    &E_5=\{(s^n(w),x^n(w,m),y^n)\in T_\epsilon^n(S,X,Y)~\text{for some}~w\neq 1\}.
\end{align}

By the covering lemma, the first term on the right side of (\ref{error:r}) vanishes as $n\rightarrow\infty$, provided that:
\begin{align}\label{3}
    R'\geq I(X;Y|S).
\end{align}

The second and third terms on the right side of (\ref{error:r}) vanish as $n\rightarrow\infty$, provided that the conditions in (\ref{1}) and (\ref{2}) are satisfied. According to the joint typicality lemma \cite{Cover}, the last term also vanishes as $n\rightarrow\infty$, provided that:
 \begin{align}\label{4}
     R+R'\leq I(S,X;Y).
 \end{align}

 Finally, by combining (\ref{1}), (\ref{2}), (\ref{3}), and (\ref{4}), and applying the Fourier-Motzkin elimination technique and the data processing inequality \cite{Cover}, we demonstrate that if the semantic rate and the message rate expressing each semantic $(R,R')$ satisfy (\ref{rates}), the average error probability $P_e^{(n)}$ asymptotically approaches zero as $n\rightarrow\infty$.
\end{proof}
\section{Conclusion}
In this paper, we presented a probabilistic model for semantic communication grounded in a rigorous philosophical conception of information, moving beyond intuition-driven models based solely on human language. By leveraging the notion of data as constraining affordances mediated by \ac{LoA}, our approach allows for a domain-independent characterization of semantic content applicable to a wide range of data. To systematically address the complexity inherent in semantic communication, we decomposed the general problem into multiple sub-problems. Focusing specifically on the subproblem where semantic ambiguity arises purely from physical channel noise, we derived a lower bound on the semantic capacity and showed that it generalizes Shannon's model as a special case. Moreover, we demonstrated that semantic communication in this setting can achieve rates for transmissible messages exceeding Shannon’s capacity. These insights not only deepen our theoretical understanding of semantic communication but also lay the groundwork for addressing more general scenarios in future work.


\begin{thebibliography}{9}

\bibitem{Shannon}
C.~E.~Shannon,
  \emph{"A mathematical theory of communication,"} 
  Bell Syst. Tech. J., vol. 27, no. 3, pp. 379-423, 1948. 

\bibitem{Weaver}
C.~E.~Shannon, and W.~Weaver,
  \emph{"The Mathematical Theory of Communication,"} 
  Urbana, IL, USA:Univ. Illinois Press, 1949. 

\bibitem{Floridi:1}
L.~Floridi,
  \emph{"Is Information Meaningful Data?,"} 
  Philosophy and Phenomenological Research 70(2), 351–370 (2005).
  
\bibitem{Sommaruga}
L.~Floridi,
  \emph{"Philosophical Conceptions of Information. In: G. Sommaruga. (eds) Formal Theories of Information.,"} 
  Lecture Notes in Computer Science, vol 5363. Springer, Berlin, Heidelberg, 2009.

\bibitem{Carnap}
 R. Carnap, and Y. Bar-Hillel,
  \emph{"An outline of a theory of semantic information,"} 
  Oct. 1952.

\bibitem{Floridi:2}
 L. Floridi,
  \emph{"Outline of a theory of strongly semantic information,"} 
  Minds Mach. 14, 2 (2004), 197–221.

\bibitem{Gunduz}
 D. Gündüz et al,
  \emph{"Beyond Transmitting Bits: Context, Semantics, and Task-Oriented Communications,"} 
  IEEE Journal on Selected Areas in Communications, vol. 41, no. 1, pp. 5-41, Jan. 2023, doi: 10.1109/JSAC.2022.3223408.

\bibitem{Qin}
 Z. Qin et al,
  \emph{"Semantic communications: Principles and challenges,"} 
   arXiv 2021, arXiv:2201.01389.

\bibitem{Shi}
 G. Shi et al,
  \emph{"From semantic communication to semantic-aware networking: Model, architecture, 
   and open problems,"} 
   IEEE Commun. Mag. 2021, 59, 44–50.

\bibitem{Hu}
 Q. Hu et al,
  \emph{"Robust semantic communications with masked VQ-VAE enabled codebook,"} 
   IEEE Trans. Wirel. Commun. 2023, 22, 8707–8722.

\bibitem{Bao}
 J. Bao et al,
  \emph{"Towards a theory of semantic communication,"} 
   In Proceedings of the 2011 IEEE Network Science Workshop, West Point, NY, USA, 22–24 June 2011; IEEE: New York, NY, USA, 2011; pp. 110–117.

\bibitem{Ma}
 S. Ma et al,
  \emph{"A Theory for Semantic Communications,"} 
   arXiv 2023, arXiv:2303.05181.

\bibitem{Shao}
 Y. Shao et al,
  \emph{"A Theory of Semantic Communication,"} 
   IEEE Trans. Mobile Computing. 2024, vol. 23, no. 12, 12211-12228 .

\bibitem{Cover}
 T. M. Cover, and J. A. Thomas,
  \emph{"Elements of Information Theory, second edition,"} 
   New York: Wiley, 2006.
  
\end{thebibliography}
\end{document}